# Future Energy Consumption Prediction Based on Grey Forecast Model

Yuan Zeng, Miao Luo and Yuzhong Liu
School of Mathematics & School of Data and Computer Science
Sun Yat-Sen University
Guangzhou, China
{zengy73, luom27, liuyzh9}@mail2.sysu.edu.cn

## Abstract

Energy issue is a hot issue in the world today. In this paper, we build a series of models to analyze the energy efficiency in four states, namely Arizona, California, New Mexico and Texas. And we predict their energy consumption in the future.

First, we classify energy according to two criteria and use MATLAB to analyze trends in energy usage in different states. We use multiple types of charts to display the relevant data.

Second, we assess the energy efficiency of each state. By looking up scholarly resource, we decide to use the DEA model to evaluate energy efficiency. During the 50 years from 1960 to 2009, the average energy efficiency of Arizona, California, New Mexico and Texas are 0.997, 0.984, 0.983 and 0.987 respectively. The average efficiency of Arizona is the highest. As for cleaner renewable energy, their pure technology efficiencies are 0.982, 0.984, 0.954 and 0.982 respectively. The overall efficiency of four states is high and the differences between them are small. Meanwhile, Arizona and Texas have the highest efficiency. In the energy efficiency analysis in 2009, by comparing the scale efficiency, we conclude that Texas present the "best" profile.

Third, we use gray forecast model to predict the future consumption of different types of energy sources in each state. In Arizona, for example, we expect energy consumption would reach 2617441.776 billion btu in 2025 and 5353152.658 billion btu in 2050. The fitting results of cleaner energy and renewable energy use are more accurate and will increase in a steady trend, while non-cleaner energy use would decrease.

Finally, we test the DEA model and the gray forecast model and make some improvements in view of their shortcomings, which make the evaluation and prediction results more accurate. We also provide perspectives for further research.

# Contents

# 1. Introduction

## 1.1 Background

Energy is the foundation for economic development and social progress and an important factor affecting the residential environment of mankind. In many aspects of development of society, such as industrialization, agricultural modernization and urbanization, energy plays a decisive role. Energy production and usage are a major portion of any economy. Thus, it is very important to analyze the energy profile and predict energy consumption structure of the future. In the United States, energy policy are decentralized to the state level.

## 1.2 Our Work

We not only analyze the performance of the energy efficiency of each state, and the usage of cleaner, renewable energy sources, but also predict the energy profile of each state in the future.

Our work can be divided into several parts:

- Analyze the energy profile for each of the four states based on the data provided.

- Build an evaluation model to describe the evolution of the energy profile of each of the four states from 1960-2009. Then use that model to determine the "best" profile for use of cleaner, renewable energy in 2009.

- Build a prediction model to predict the energy profile of each state for 2025 and 2050. Based on this prediction and the previous model, find the renewable energy usage targets for 2025 and 2050.

# 2. Assumptions and Parameters

## 2.1 Assumptions and Justifications

- Each variable is independent to each other
- Ignore some minor energy losses
- Sufficient energy sources are available
- Adopt VRS variable model reward condition in DEA model

- The world is actively using all technologies and energy sources to replace fossil fuels
- Predict energy use only take a few key factors, other factors are temporarily ignored

## 2.2 Parameters

| NOTATIONS | DEFINITION |
|---|---|
| $DMU_j(j=1,2,...,n)$ | Independent decision-making unit. |
| $x_w$ | input |
| $y_q$ | output |
| $\theta$ | Technical efficiency value |
| $\lambda_j$ | Mix proportion |
| $s_i^-, s_r^+$ | slack variable |
| $X^{(0)}$ | the original data sequence |
| $X^{(1)}$ | The corresponding generated data sequence |
| $Z^{(1)}$ | The mean generating sequence |
| A,B | the parameter that needs to be solved |
| BBTU | Billion btu |

# 3. Analy

# 4. DEA Model

## 4.1 Energy Profile

In this section, we describe the energy profile in two methods respectively. In the first approach, we divide the energy-consuming sectors into four [2]:
- Industrial
- Transportation
- Commercial
- Residential

Then we draw up line charts from the data of the past 50 years for each of the four departments in each state and analyze it.

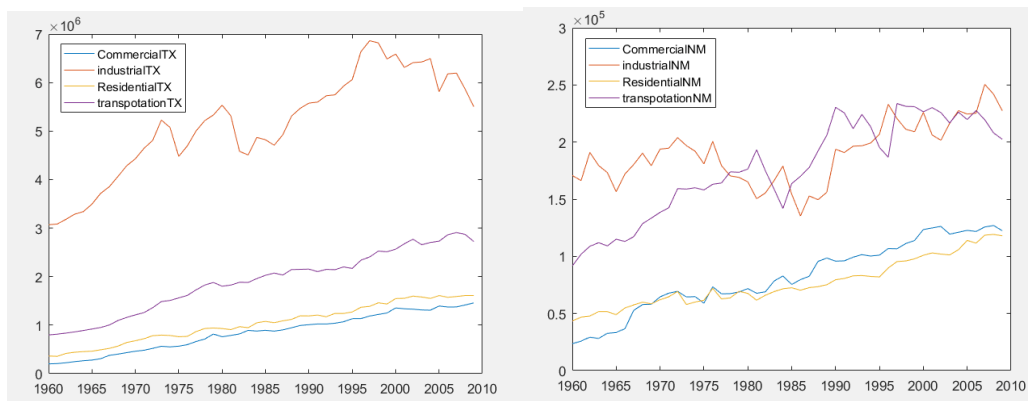

Figure 4.1

As we can see in the graph above, the energy consumption of the transportation sector in Texas has been steadily increasing over the past 50 years and nearly doubled from the beginning of 1960[3]. What's more, industrial energy consumption has increased and reached its peak in 2009. Residential and commercial energy consumption have remained stable over the past 50 years.

We can see in the figure 4.2 above that energy consumption of transportation sector in New Mexico surpassed that of industrial and become the most energy-consuming sector. Both industrial and transportation sectors experienced great fluctuation in energy consumption.  Residential and commercial energy consumption have remained stable over the past 50 years.

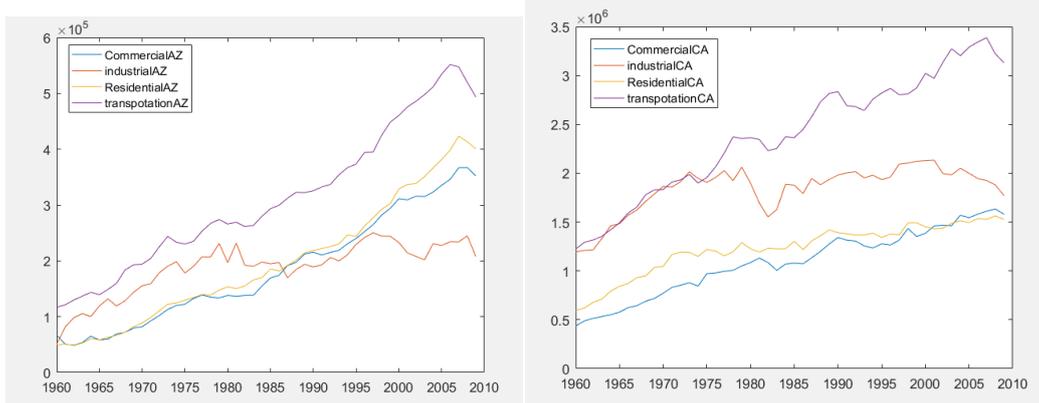

Figure 4.2

The consumption of energy of all sectors in Arizona has been increased over time, with transportation accounting for the major portion, with little difference between other sectors. However, due to the cleaner energy came into market, transportation consumption has been declined recently.

The energy consumption of four sectors in California has been risen over time, with transportation and industrial consumption accounting for the main share [4]. However, due to the implementation of policies such as environmental protection policy, the upward trend of consumption in the transportation and industrial sectors began to gradually change into a downward and steady trend.

Since the economy developed rapidly, there are discrepancies in the purchasing power of dollars [5]. In addition, the basic units of different energy sources are different. Therefore, we choose BTU as the uniform measure.

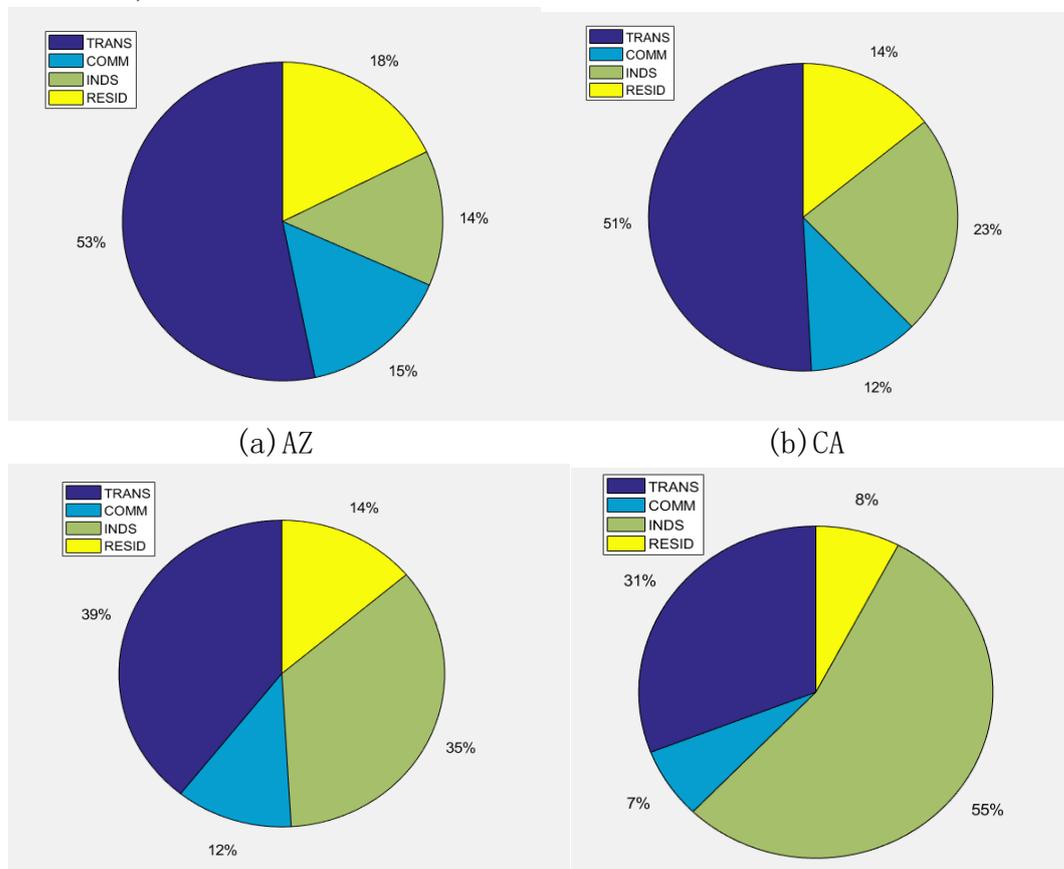

(a) AZ     (b) CA

(c) NM　　　　　　　　　　　　(d) TX

Figure 4.3

The above are pie charts of the energy consumption of the four major sectors in the four states in 2009. We can conclude that in Arizona, California and New Mexico, transportation sector consumed most of the energy[6]. While the other sectors accounted for little difference. In Texas, however, residential sector consumes a major portion. Because of the different economic development in each state, composition ratio differs greatly.

In the latter process, we divide the energy into cleaner renewable, cleaner non-renewable, non-cleaner non-renewable and non-cleaner renewable energy sources. And then draw up line charts with the consumption of these four energy sources in each state in the past 50 years.

The following are the definitions of different type energy:

| ENERGY TYPE | Energy name |
| --- | --- |
| CLEANER RENEWABLE ENERGY (CRN) | Electricity, geothermal energy, ethanol and so on |
| CLEANER NON-RENEWABLE ENERGY (CNRN) | natural gas |
| NON-CLEANER NON-RENEWABLE ENERGY (NCNRN) | Wood, waste and so on |
| NON-CLEANER RENEWABLE ENERGY (NCRN) | Coal, oil (including asphalt, gasoline, lubricants, etc.) and so on |

First, we divide the energy in the data into the four categories above.

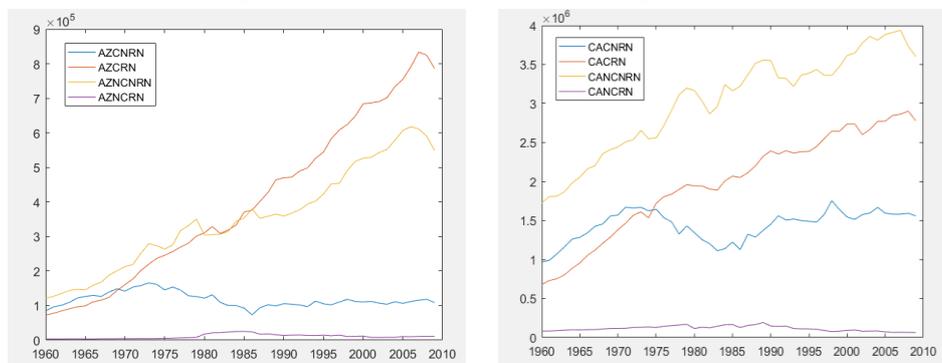

Figure 4.4

As can be seen from the figure 4.4 above, AZ has increased their use of cleaner renewable, non-renewable and non-cleaner and renewable energy sources increasing at a much larger rate, about 1.3 times. The other two uses of energy are maintained at a relatively stable level.

In California, the use of cleaner renewable and non-cleaner non-renewable energy both show a steady increasing rise but they had a stable difference; the utilization of cleaner non-renewable energy sources also showed a slight upward trend but it had an obvious fluctuation between 1960 and 1985; the use of non-cleaner non-renewable energy sources almost stabilized at a smaller value.

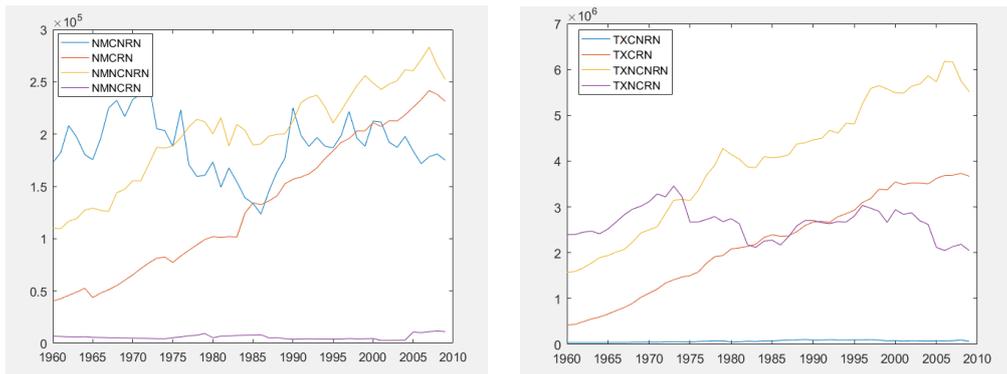

Figure 4.5

In New Mexico, the use of cleaner renewable and non-renewable energy sources also show a steady increasing rise and the use of cleaner renewable energy will be almost catching up with the pollution of non-renewable energy sources in the latest future[7]. The use of cleaner non-renewable energy was demonstrated A greater volatility; the use of renewable sources of energy such as wood is also kept at a low level.

In TX, however, unlike the previous three states, the state's use of cleaner energy contaminated renewable energy and cleaner non-renewable energy are different compared with those of the previous states, but the other two energy uses witness a steady rise[8]. Meanwhile, the use of pollution of non-renewable energy are more stable than the use of cleaner renewable energy.

## 4.2 Introduction of DEA Model

The DEA method uses a mathematical programming model to evaluate the same type of decision making unit (DMU) A nonparametric statistical method of whether the unit is technology effective or not. Its theoretical basis is derived from the envelope thought put forward by Farrell. The first Model of DEA method (That is, the CCR-DEA model) was first proposed, in 1978, by three famous operational researchers, Charnes, Cooper and Rhodes. After more than 30 years of development, the theory of DEA was becoming more and more perfect and the method was becoming more and more mature. At present, it has become an important analytical tool in management science and evaluation technology and it has been successfully applied in many research fields[9].

Assume there are n independent decision-making units $DMU_j (j = 1, 2, ..., n)$, each of which has w kinds of resource input $x_w$ and $q$ kinds of output $y_q$, the inputs and outputs of the j decision-making unit $DMU_j$ are as follows:

$$x_j = \left( x_{1j}, x_{2j}, ..., x_{wj} \right)^T > 0$$

$$y_j = \left(y_{1j}, y_{2j}, ..., y_{qj}\right)^T > 0$$

Assume we will evaluate the $j_0$ decision-making unit, which is recorded as $DMU_0$, according to the linear programming duality theory, the model is:

$$\begin{cases} \min \theta \\ s.t. \sum_{j=1}^{n} \lambda_j x_j + s_i^- = \theta x_0, i = 1, 2, ...w \\ \sum_{j=1}^{n} \lambda_j y_j - s_r^+ = y_0, r = 1, 2, ..., q \\ \sum_{j=1}^{n} \lambda_j = 1, \lambda_j \geq 0 \\ s_i^- \geq 0 \\ s_r^+ \geq 0 \end{cases} \quad (I)$$

$\theta$ is the technology efficiency of the decision-making unit $DMU_0$ ($\theta \leq 1$), $\lambda_j$ is, compared with $DMU_0$, the combination ratio of the jth decision unit, $DMU_j$, of the reconstruction of a combination of $DMU$. $s_i^-, s_r^+$ are $w$ variables of input and $q$ kinds of output respectively.

The meanings of the model are as follows[10]:

① If $\theta = 1$ and $s_i^- = 0, s_r^+ = 0$, it shows that the decision-making unit is the effective unit of DEA and at the same time achieves the effective technology and scale, the efficiency of the formation of its frontier will be the same scale returns;

② If $\theta < 1$ and $s_i^- \neq 0, s_r^+ \neq 0$, that means the decision-making unit is an invalid unit, the technology or the scale may be invalid.

③If the decision-making unit is invalid, it can be improved according to the projection of the decision-making unit on the efficiency frontier. Through the $s_i^-, s_r^+$ level of decision-making unit can determine the input redundancy and output deficit, that is, for the input $x_0$ of the decision-making unit can be reduced $s_i^-$ to maintain the same level of output $y_0$, or inputs $x_0$ can be unchanged, the output $y_0$ can be increased to $s_r^+$. This is also the fundamental basis for this study to determine the potential of energy conservation and emission reduction.

This model breaks through the limitations of the constant return scale (CRS), which can distinguish pure technology efficiency (PTE) and scale efficiency (SE) under the condition of variable return (VRS). The pure technical efficiency measures the distance between the DMU and the efficiency frontier when the scale returns variable. The scale efficiency refers to the efficiency frontier with constant returns to scale and the efficiency frontier with variable returns to scale the distance between the faces. Which is measured in mathematical expressions is that technical efficiency = Pure technical efficiency × Scale efficiency.

The scale efficiency of decision-making unit can determine its scale income: if the scale efficiency is $SE = 1$, and the decision-making unit for the return of scale keeps same, showing that when all the input elements of the decision-making unit increase in the same proportion, it will receive the same proportion of output; If the efficiency of scale is $SE < 1$, there are two possibilities, if $\sum_{j=1}^{n} \lambda_j < 1$, the decision-making unit for the scale of return will increase, shows that if the decision unit investment increase in the same proportion, it will get a greater proportion of the return; if $\sum_{j=1}^{n} \lambda_j > 1$, the decision unit for the scale of income decrease, indicating that after the increasing input of the decision-making unit, the rate of the output growth will be less than the proportion of increase.

The difference between pure technical efficiency and scale efficiency is shown in the following figure. Decision-making unit I and G both have a kind of input X and a kind of output Y. CRS is the efficiency frontier with constant returns to scale and VRS is the efficiency frontier with variable returns to scale. Both decision-making units I and G are technical Invalid units, VRS is more compact than the CRS based on the data envelopment of the CRS and the VRS. Therefore, the efficiency value obtained by the CRS (That is, the technical efficiency TE) is less than or equal to the efficiency value obtained by the VRS (That is, the pure technical efficiency PTE). The product of the two is the scale efficiency SE. Take decision unit I as an example, its technical efficiency TE = HA / HI, pure technical efficiency PTE = HB / HI, and scale efficiency SE = HA / HB.

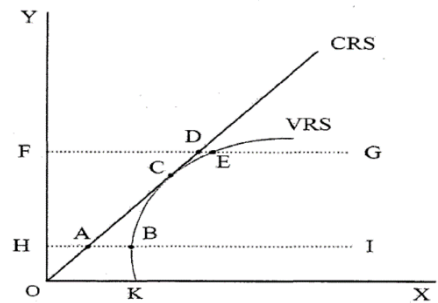

Figure 4.5

By examining the curves of CRS and VRS, we can also find that when the decision-making units are produced on the CRS curve, they are in constant returns in scale. When the decision-making unit is produced on the VRS curve, it is in the state of variable returns in scale, and it will increase in scale on the KC segment, and decreases in scale on the CE segment.

The equation $(I)$ is based on input-oriented DEA, and its corresponding output-oriented DEA model equation $(O)$ is:

$$\begin{cases} \max \eta \\ s.t. \sum_{j=1}^{n} \lambda_j y_j - s_r^+ = \eta y_0, r = 1, 2, \ldots q \\ \sum_{j=1}^{n} \lambda_j x_j + s_i^- = x_0, i = 1, 2, \ldots, w \\ \sum_{j=1}^{n} \lambda_j = 1, \lambda_j \geq 0 \\ s_i^- \geq 0 \\ s_r^+ \geq 0 \end{cases} \quad (O)$$

In the point of the economic, the equation $(I)$ means that if the decision-making unit $\theta < 1$, shows that the decision-making unit can obtain the same output with less investment, that is, under certain levels of output, all kinds of inputs can be reduced by the same proportion $\theta$ to the minimum. Equation $(O)$ means that if the decision-making unit $\eta < 1$, the decision-making unit can use the same input to obtain greater output, that is, under certain input conditions, the output can be increased in accordance with the same proportion $\eta$ to the maximum.

The efficiency frontier between the DEA model and the DEA model is the same, but the calculation results may have some differences. However, Coelli thinks this different guidance method is much smaller than that caused by different calculation methods Differences. In most current efficiency studies, we choose input-oriented DEA models based primarily on the fact that decision-making units make it easier to control inputs relative to the output [11].

## 4.3 The Application of DEA Model

We use the DEA model into our states' energy profile. And we can get the following details.

Table 1: DEA evaluation values of classification energy use efficiency of AZ in 1960-2009

| Year | Output indicator | | | | Input indicator | | | Valid value |
|---|---|---|---|---|---|---|---|---|
| | The usage amount of clean renewable(BBTU) | The usage amount of clean non-renewable(BBTU) | The usage amount of non-clean renewable(BBTU) | The usage amount of non-clean non-renewable(BBTU) | Population (ten thousand people) | Total revenue (handred million dollor) | Total end-use energy average price (Dollars per billion Btu) | |
| 1960 | 0 | 0 | 0 | 0 | 0 | 0 | 0 | 1 |
| 1961 | 0 | 0 | 0 | 0 | 0 | 0 | 0 | 1 |
| 1962 | 0 | 979.076 | 283.116 | 136.287 | 0 | 0 | 0 | 0.985 |
| 1963 | 0 | 0 | 0 | 0 | 0 | 0 | 0 | 1 |
| 1964 | 0 | 0 | 0 | 0 | 0 | 0 | 0 | 1 |
| 1965 | 0 | 0 | 0 | 0 | 0 | 0 | 0 | 1 |
| 1966 | 0 | 0 | 0 | 0 | 0 | 0 | 0 | 1 |
| 1967 | 0 | 0 | 0 | 0 | 0 | 0 | 191.316 | 0.991 |
| 1968 | 0 | 0 | 0 | 0 | 0 | 0 | 0 | 1 |
| 1969 | 0 | 0 | 0 | 0 | 0 | 0 | 0 | 1 |
| 1970 | 0 | 0 | 0 | 0 | 0 | 0 | 0 | 1 |
| 1971 | 0 | 0 | 77.714 | 11579.494 | 0 | 1.451 | 195.163 | 0.997 |
| 1972 | 0 | 0 | 0 | 0 | 0 | 0 | 0 | 1 |
| 1973 | 0 | 0 | 0 | 0 | 0 | 0 | 0 | 1 |
| 1974 | 0 | 0 | 0 | 0 | 0 | 0 | 0 | 1 |
| 1975 | 0 | 0 | 0 | 0 | 0 | 0 | 0 | 1 |
| 1976 | 0 | 0 | 0 | 0 | 0 | 0 | 0 | 1 |
| 1977 | 0 | 0 | 0 | 0 | 0 | 0 | 0 | 1 |
| 1978 | 0 | 0 | 0 | 0 | 0 | 0 | 0 | 1 |
| 1979 | 0 | 0 | 0 | 0 | 0 | 0 | 0 | 1 |
| 1980 | 0 | 14002.296 | 0 | 710.603 | 0 | 5.525 | 1219.431 | 0.992 |
| 1981 | 0 | 0 | 0 | 0 | 0 | 0 | 0 | 1 |
| 1982 | 0 | 0 | 0 | 0 | 0 | 0 | 0 | 1 |
| 1983 | 0 | 0 | 0 | 0 | 0 | 0 | 0 | 1 |
| 1984 | 0 | 0 | 0 | 0 | 0 | 0 | 0 | 1 |
| 1985 | 0 | 0 | 0 | 0 | 0 | 0 | 0 | 1 |
| 1986 | 0 | 0 | 0 | 0 | 0 | 0 | 0 | 1 |
| 1987 | 0 | 12549.122 | 0 | 0 | 0 | 14.006 | 0 | 0.958 |
| 1988 | 0 | 0 | 0 | 4846.367 | 0 | 0 | 0 | 0.989 |
| 1989 | 0 | 0 | 0 | 0 | 0 | 0 | 0 | 1 |
| 1990 | 0 | 0 | 0 | 0 | 0 | 0 | 0 | 1 |
| 1991 | 0 | 0 | 0 | 0 | 0 | 0 | 0 | 1 |
| 1992 | 0 | 312.492 | 0 | 1670.406 | 4.854 | 0 | 369.705 | 0.989 |
| 1993 | 0 | 8260.747 | 998.922 | 0 | 13.955 | 0 | 510.769 | 0.997 |
| 1994 | 0 | 0 | 0 | 0 | 0 | 0 | 0 | 1 |
| 1995 | 0 | 5411.766 | 0 | 0 | 11.42 | 0 | 0 | 0.992 |
| 1996 | 0 | 9976.205 | 685.391 | 0 | 12.915 | 0 | 69.623 | 0.967 |
| 1997 | 0 | 0 | 0 | 0 | 0 | 0 | 0 | 1 |
| 1998 | 0 | 0 | 0 | 0 | 0 | 0 | 0 | 1 |
| 1999 | 0 | 0 | 0 | 0 | 0 | 0 | 0 | 1 |
| 2000 | 0 | 0 | 0 | 0 | 0 | 0 | 0 | 1 |
| 2001 | 0 | 0 | 0 | 0 | 0 | 0 | 0 | 1 |
| 2002 | 0 | 0 | 0 | 0 | 0 | 0 | 0 | 1 |
| 2003 | 0 | 6044.084 | 996.736 | 0 | 1.464 | 0 | 0 | 0.986 |
| 2004 | 0 | 0 | 0 | 0 | 0 | 0 | 0 | 1 |
| 2005 | 0 | 0 | 0 | 0 | 0 | 0 | 0 | 1 |
| 2006 | 0 | 0 | 0 | 0 | 0 | 0 | 0 | 1 |
| 2007 | 0 | 0 | 0 | 0 | 0 | 0 | 0 | 1 |
| 2008 | 0 | 0 | 0 | 0 | 0 | 0 | 0 | 1 |
| 2009 | 0 | 5791.107 | 0 | 38506.552 | 36.346 | 22.026 | 0 | 0.986 |

Note: The unit "bbtu" in the table means "billion btu". The valid values in the table are the technical efficiencies when considering scale benefits (pure technical efficiency).

Overall energy input and output efficiency of each state: During the period from 1960 to 2009, the average energy efficiencies of AZ, CA, NM and TX are 0.997, 0.984, 0.983 and 0.987. The overall level of energy use efficiency is high but their difference is not significant, and AZ has the highest average efficiency.

During the 50 years, AZ has 37 years whose efficiency was 1, in other words, there are 37 years, which has no room for reducing inputs with output fixed. In technologically effective year, the average efficiency of Az is 0.957. At the same time, DEA was ineffective in 13 years, taking 1971 as an example. In the case of fixed output, there was a redundancy of 145.1 million total revenues and an average energy price of 195.163 dollars per bbtu in 1971. And in the case of fixed investment, it can increase the usage amount of contaminated renewable energy as 77.71 bbtu, the amount of non-cleaner non-renewable energy use as 11579.964 bbtu. See Table 1 for details.

For the other three states, the form cannot be given due to space limitations, but we can analyze it similarly.

In the past 50 years, CA has been inactive for 20 years, of which 1982 has the lowest efficiency of 0.918, which means the input can be reduced by a large margin.

In the past 50 years of NM, DEA was ineffective in 21 years, of which 1983 had the lowest efficiency of 0.884.

In the past 50 years of TX, there were 25 years with an efficiency of 1, compared with the same number of years in which DEA was ineffective.

Overall cleaner renewable energy input and output efficiency of each state: We set the output indicators of the DEA model only as the usage amount of cleaner renewable energy, and input indicators are as same as before, then we use DEA model again to measure the efficiency.

According to the result, in 1960-2009, the pure technology efficiencies of AZ, CA, NM and TX were 0.982, 0.973, 0.954 and 0.982. As same as before, the overall efficiency was high and the difference was not significant. Among them, AZ and TX had the highest efficiency. However, the average efficiency is obviously lower than the average efficiency of the previous case. It can be seen that there is still much space for improvement of their efficiency.

According to the table 2, AZ only had 20 years whose efficiency was 1 in 50 years. Through data analysis, we can see that if we keep the output unchanged and adjust the quantity and structure of input indicators, there will be a total redundancy of 650,500 people and 28.78 billion US dollars in 50 years.

For the remaining three states, from data analysis, the amount of years whose efficiency was 1 in CA was only 16, that in NM was only 13, and that in TX was only 18. There were also too much input redundancy.

Table 2: DEA evaluation values of cleaner renewable energy use efficiency of AZ in 1960-2009

| Year | Output indicator | Input indicator | | | Valid value |
|---|---|---|---|---|---|
| | The usage amount of clean renewable(BBTU) | Population (ten thousand people) | Total revenue (handred million dollor) | Total end-use energy average price (Dollars per billion Btu) | |
| 1960 | 0 | 0 | 0 | 0 | 1 |
| 1961 | 0 | 0 | 0 | 0 | 0.979 |
| 1962 | 0 | 0 | 0 | 0 | 0.985 |
| 1963 | 0 | 0 | 0 | 0 | 1 |
| 1964 | 0 | 0 | 0 | 0 | 1 |
| 1965 | 0 | 0 | 0 | 0 | 1 |
| 1966 | 0 | 0 | 0 | 0 | 1 |
| 1967 | 0 | 0 | 0 | 0 | 0.95 |
| 1968 | 0 | 0 | 0 | 469.385 | 0.945 |
| 1969 | 0 | 0 | 0.393 | 655.372 | 0.972 |
| 1970 | 0 | 0 | 0 | 0 | 1 |
| 1971 | 0 | 0 | 0.308 | 0 | 0.987 |
| 1972 | 0 | 0 | 0 | 0 | 1 |
| 1973 | 0 | 0 | 0 | 0 | 1 |
| 1974 | 0 | 0 | 2.287 | 0 | 1 |
| 1975 | 0 | 0 | 0 | 0 | 1 |
| 1976 | 0 | 0 | 0 | 0 | 1 |
| 1977 | 0 | 0 | 0 | 0 | 1 |
| 1978 | 0 | 0 | 16.641 | 157.991 | 0.991 |
| 1979 | 0 | 0 | 13.121 | 1542.207 | 0.994 |
| 1980 | 0 | 0 | 12.892 | 4921.071 | 0.986 |
| 1981 | 0 | 0 | 4.253 | 5127.028 | 0.993 |
| 1982 | 0 | 0 | 7.733 | 4047.177 | 0.922 |
| 1983 | 0 | 0 | 11.025 | 2469.019 | 0.919 |
| 1984 | 0 | 0 | 34.767 | 1763.669 | 0.918 |
| 1985 | 0 | 0 | 23.332 | 2152.713 | 0.957 |
| 1986 | 0 | 0 | 40.297 | 0 | 0.932 |
| 1987 | 0 | 0 | 33.013 | 239.601 | 0.943 |
| 1988 | 0 | 0 | 12.695 | 0 | 0.96 |
| 1989 | 0 | 0 | 0 | 0 | 1 |
| 1990 | 0 | 0 | 0 | 0 | 1 |
| 1991 | 0 | 0 | 0 | 0 | 1 |
| 1992 | 0 | 0.406 | 0 | 0 | 0.985 |
| 1993 | 0 | 6.003 | 0 | 0 | 0.985 |
| 1994 | 0 | 5.272 | 0 | 0 | 0.994 |
| 1995 | 0 | 7.408 | 0 | 0 | 0.988 |
| 1996 | 0 | 8.606 | 0 | 0 | 0.958 |
| 1997 | 0 | 0 | 0 | 0 | 1 |
| 1998 | 0 | 0.587 | 0 | 0 | 0.998 |
| 1999 | 0 | 0 | 0 | 0 | 1 |
| 2000 | 0 | 0 | 0 | 0 | 1 |
| 2001 | 0 | 0.635 | 0 | 0 | 0.998 |
| 2002 | 0 | 0 | 0 | 0 | 1 |
| 2003 | 0 | 0.215 | 0 | 0 | 0.978 |
| 2004 | 0 | 0 | 0.187 | 0 | 0.988 |
| 2005 | 0 | 0 | 26.213 | 0 | 0.967 |
| 2006 | 0 | 0 | 48.661 | 0 | 0.977 |
| 2007 | 0 | 0 | 0 | 0 | 1 |
| 2008 | 0 | 8.617 | 0 | 2719.2 | 0.989 |
| 2009 | 0 | 27.318 | 0 | 0 | 0.974 |

Note: The valid values in the table are the technical efficiencies when considering scale benefits (pure technical efficiency).

**Difference analysis:** After comparing input data from each state, we can conclude:
1. NM has the lowest efficiency in all DEA models because of the lowest per capita income.
2. Although per capita income in CA is the highest among the four states, energy efficiency is low among the four states as both population and energy prices are the highest.
3. Comparing AZ with TX, the per capita income and energy prices in AZ are almost the same as that of TX, but the population of AZ is much smaller than that of TX, so the energy efficiency of AZ is better than that of TX.

## 4.4 Improved DEA Model

In the analysis of energy efficiency in 2009, we add more data into the model so that the model could be more accurate. Therefore we get an improved model. Hence we can get more accurate assessments. The data are shown in table 3. As we can see from the analysis in table 4, the scale efficiency of the four states are increasing in order, which indicates that the scale efficiency in TX is the best, showing the "best" profile.

Table 3 Energy input and output indicators of four states in 2009

| state | Output indicator | Input indicator | | | | |
|---|---|---|---|---|---|---|
| | The usage amount of clean renewable(BBTU) | Total revenue (handred million dollor) | Total end-use energy average price (Dollars per billion Btu) | Urban population ratio | Percent of adults with more than a high school diploma | Unemployment rate |
| AZ | 785814.161 | 2140.509 | 19664.82377 | 89.8 | 9.9 | 81.0 |
| CA | 2779376.455 | 15605.990 | 18405.2811 | 95 | 11.2 | 76.8 |
| NM | 231431.806 | 662.494 | 17178.97394 | 77.4 | 7.5 | 78.9 |
| TX | 3669647.585 | 9100.360 | 15379.24605 | 84.7 | 7.6 | 75.7 |

Table 4 DEA efficiency of cleaner renewable energy in 2009

| state | crste | vrste | scale | |
|---|---|---|---|---|
| AZ | 0.91 | 0.967 | 0.941 | irs |
| CA | 0.747 | 0.986 | 0.757 | irs |
| NM | 0.866 | 1 | 0.866 | irs |
| TX | 1 | 1 | 1 | - |

Note: Crste is the technical efficiency (aggregate efficiency) when the scale income are not taken into account. Vrste is the technical efficiency (purely technical efficiency) when considering the scale income. Scale is the scale efficiency when considering the scale income. Irs, drs, - are the returns of scale increasing, decreasing, the same.

# 5 Gray forecast model

## 5.1 Series Prediction GM (1,1) Model

GM model derived from the differential equation of gray system theory, G represents gray, M represents model, and Gm (1,1) represents first order, one variable differential equation model[11].

The modeling process and mechanism of Gm (1,1) are as follows:
Note the original data sequence $X^{(0)}$ as Non-negative sequence

$$X^{(0)} = \{x^{(0)}(1), x^{(0)}(2), x^{(0)}(3), ..., x^{(0)}(n)\}$$

among them, $x^{(0)}(k) \geq 0, k = 1, 2, \cdots, n$

The corresponding generated data sequence is $X^{(1)}$

$$X^{(1)} = \{x^{(1)}(1), x^{(1)}(2), x^{(1)}(3),...,x^{(1)}(n)\}$$

among them, $x^{(1)}(k) = \sum_{i=1}^{k} x^{(0)}(i), k = 1,2,\cdots,n$

$Z^{(1)}$ is the nearest mean generating sequence for $X^{(1)}$

$$Z^{(1)} = \{z^{(1)}(1), z^{(1)}(2), \cdots, z^{(1)}(n)\}$$

among them

$$Z^{(1)}(k) = 0.5x^{(1)}(k) + 0.5x^{(1)}(k-1), k = 1,2,\cdots n$$

Say $x^{(0)}(k) + az^{(1)}(k) = b$ is the Gm (1,1) model, where $a$, b are the parameter that needs to be solved by modeling, if $a = (a,b)^T$ is the parameter column, and

$$Y = \begin{bmatrix} x^{(0)}(2) \\ x^{(0)}(3) \\ \vdots \\ x^{(0)}(n) \end{bmatrix}, \quad B = \begin{bmatrix} -z^{(1)}(2) & 1 \\ -z^{(1)}(3) & 1 \\ -z^{(1)}(4) & 1 \\ -z^{(1)}(5) & \end{bmatrix}$$

Then find the solution of the differential equation $x^{(0)}(k) + az^{(1)}(k) = b$ which satisfies

$$\hat{a} = (B^T B)^{-1} B^T Y$$

Call $\dfrac{dx^{(1)}}{dt} + ax^{(1)} = b$ as the gray differential equation, $x^{(0)}(k) + az^{(1)}(k) = b$ as the albino equation, also known as shadow equation.

As mentioned above, then we have:

1. The solution, time response function, of albino equation $\dfrac{dx^{(1)}}{dt} + ax^{(1)} = b$ is

$$\hat{x}^{(1)}(t) = (x^{(1)}(0) - \frac{b}{a})e^{-at} + \frac{b}{a}$$

2. The time response sequence of Gm (1,1) gray differential equation $x^{(0)}(k) + az^{(1)}(k) = b$ is

$$\hat{x}^{(1)}(k+1) = (x^{(1)}(0) - \frac{b}{a})e^{-ak} + \frac{b}{a}, k = 1,2,\cdots,n$$

3. Assume $x^{(1)}(0) = x^{(0)}(1)$, then

$$\hat{x}^{(1)}(k+1) = (x^{(0)}(1) - \frac{b}{a})e^{-ak} + \frac{b}{a}, k = 1,2,\cdots,n$$

4. Reducing value

$$\hat{x}^{(0)}(k+1) = \hat{x}^{(1)}(k+1) - \hat{x}^{(1)}(k), k = 1,2,\cdots,n$$

## 5.2 Application of Gray Forecast Model

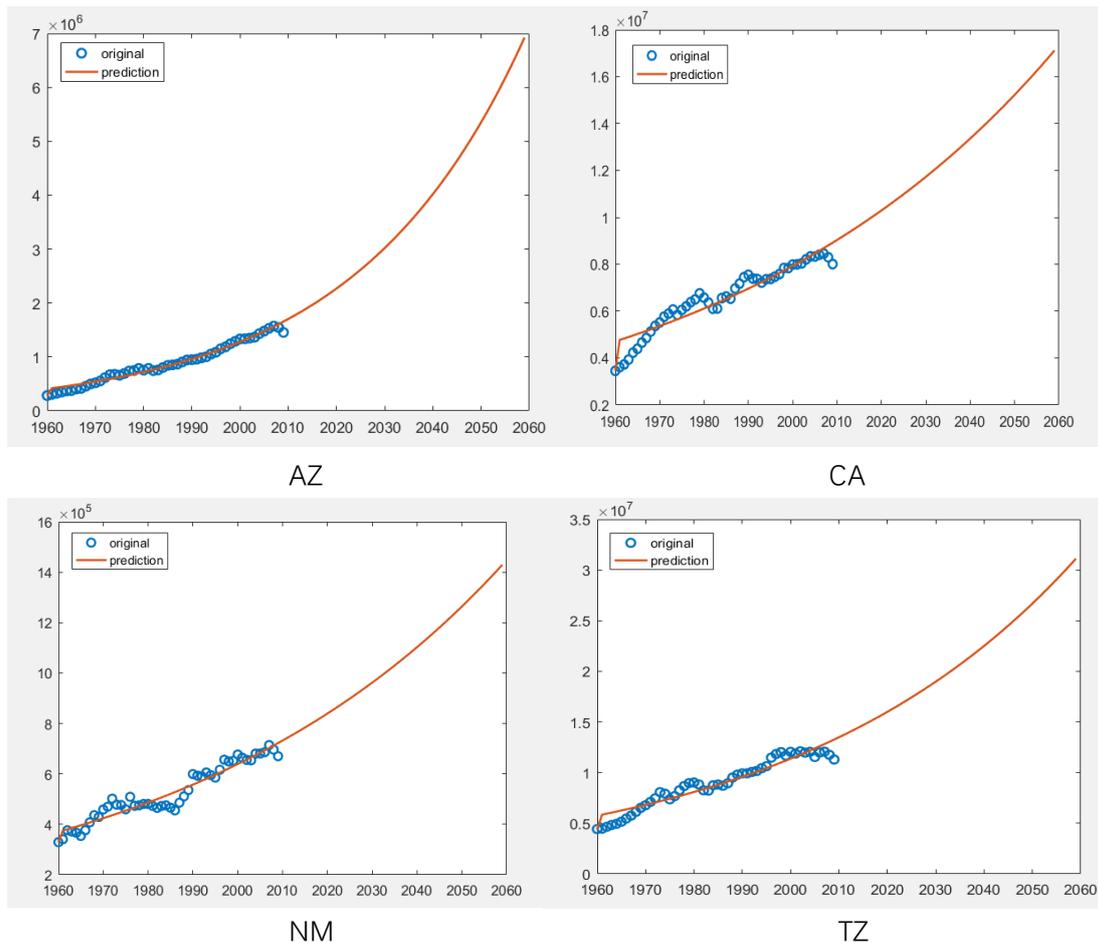

Figure 5.1

As shown in the figures 5.1, we use the gray forecast method to establish the model based on the 50-years-data of each state and forecast them in the future. We predicted that the energy consumption of AZ in 2025 will be 2617441.776 bbtu and the energy consumption in 2050 will be 5353152.658 bbtu; the energy consumption of CA in 2025 will be 1132689.36 bbtu, the energy consumption in 2050 will be 1524565.47 bbtu; the energy consumption of NM in 2025 will be 898632.0662 bbtu, and the energy consumption will be 1263979.235 bbtu in 2050; TZ energy consumption in 2025 17427523. 18 bbtu, 2050 energy consumption 26694360.43 bbtu.

From the pie chart in figure 5.2, we can see that NM has the largest proportion of cleaner energy and renewable energy, so it showed the best profile in energy mix in 2009
By the same forecasting method, we can get the predicted values of all kinds of energy sources in 2025 and 2050. The prediction of the usage amount of cleaner renewable energy sources is more accurate and will rise exponentially, and the use amount of non-cleaner energy sources will decrease [13]. Within a certain margin of error, the relevant predictive value is accurate. We use the abbreviation of the states and energy.

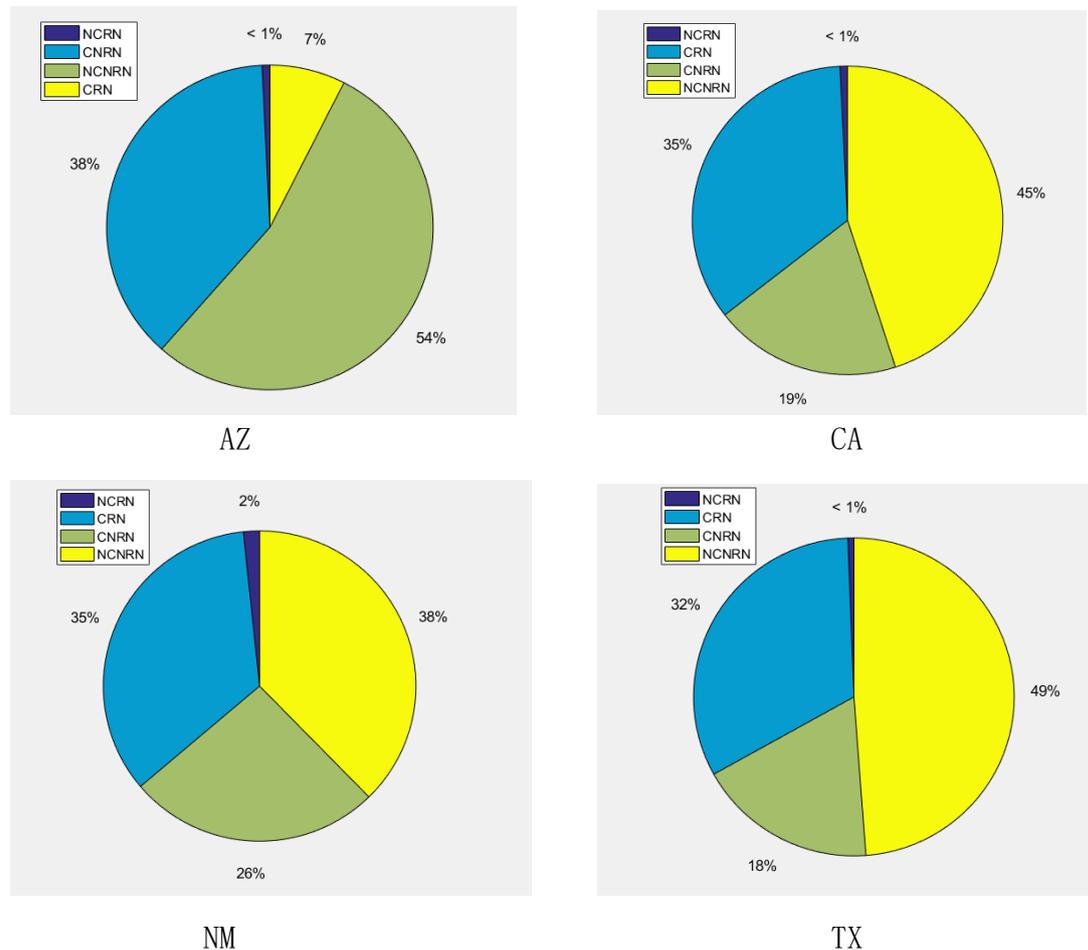

Figure 5.2

|  | **2025** | **2050** |
|---|---|---|
| **AZCRN** | 1597500 | 2005000 |
| **CACRN** | 3765750 | 4871500 |
| **NMCRN** | 311325 | 422650 |
| **TXCRN** | 5178250 | 6996500 |

Table 5

From the table 5, we can conclude the energy usage targets of the four states about the cleaner renewable energy.

## 5.3 The Suggestions of Using Cleaner Renewable Energy

Based on our model and data, we propose the following suggestions [14]:
- Establish interstates cooperation standards and take a win-win cooperation and jointly develop energy sources
- Release policies to promote the development of new industries

- Sign non-aggression agreement and execute the energy policies accurately
- Increase the usage amount of cleaner renewable energy especially solar and wind energy
- Adjust the industrial structure to maximize energy use and minimize the deterioration of renewable energy.
- Eliminate malicious waste of energy

# 6. Sensitivity Analysis and Advanced model

## 6.1 Sensitivity Analysis

We use the first 40-year-data from AZ commercial sector as a sample to predict the 1999-2009 data by using gray forecast model, then computing the errors. Specific data and prediction model are as shown below:

|  | Original data | Predicting data | error |
| --- | --- | --- | --- |
| 2000 | 311260.3 | 312200.5 | 0.003 |
| 2001 | 3093311.8 | 323202.9 | 0.045 |
| 2002 | 315794.4 | 336750.3 | 0.0664 |
| 2003 | 315230 | 350865.6 | 0.113 |
| 2004 | 323078 | 365572.5 | 0.1315 |

Table 6

To sum up, our prediction is accurate within a certain margin of error.

## 6.2 Improved Model

Because of the limited amount of energy sources, the exponential growth of energy will lead to large errors in the long-term prediction. Therefore, for the long run, we consider using linear fit model to predict the relevant energy consumption. According to the data provided, we get the following linear formula [15]:

$$Y = 5930 * year - 1.159 * 10^7$$

The projections for the energy consumption of the commercial sector in the state of AZ in 2025 and 2050 are as follows

|  | Gray forecast | Linear prediction |
| --- | --- | --- |
| **2025** | 732492 | 418250 |
| **2050** | 1828341 | 566500 |

Table 7

In the improved model, the energy consumption of 2025 will be 418250 bbtu, 566500 bbtu in 2050. Other energy sources can also be predicted by the improved model. The new forecasting result will be more realistic.

# 7. Analysis of our Model

## 7.1 Strengths

As for the DEA model, we can see its merits as follow:

- Different from the efficiency evaluation method that can only deal with single output in the past, this method can handle multiple inputs and multiple outputs, and does not need to build a production function to estimate the parameters.

- The DEA method is independent of the input-output dimension. The DEA method does not affect the final efficiency assessment due to differences in measurement units, and as long as all DMUs use the same units of measurement, the efficiency value can still be calculated.

- DEA method is used to evaluate the efficiency of a comprehensive index. This indicator represents the use efficiency of resources and is suitable for describing the data of states' total productivity, and it allows comparisons of the efficiency among DMUs.

- The DEA approach suggests an improvement for inefficient DMUs. The DEA method, by analyzing slack variables, provides further insight into the use of inefficient DMU resources and suggests directions and sizes for their inefficient resources, providing policymakers with ways to improve efficiency.

As for gray forecast model, it is easy to show its benefits.

- Less modeling information, high precision
- A variety of energy forecasts

## 7.2 Weaknesses

Of course, there are still many drawbacks of the DEA models.
- The DEA method is only a relative efficiency assessment of DMU, not an absolute efficiency assessment. Therefore, DEA cannot completely replace the traditional ratio analysis of the absolute efficiency of the analysis.
- The DEA method cannot measure negative output. The linear model assumption simplifies DEA analysis, but output is just the prerequisite for solving linear programming. If the output is negative, it cannot be measured by this method.

- The choice of inputs and outputs in the DEA method has a decisive effect on the efficiency assessment. Improper selection of inputs and outputs can affect the shape and location of the front of the production, thereby affecting the accuracy of the efficiency assessment.

As for gray forecast method, we have to admit its weaknesses.

- Regardless of the inherent mechanism of the system, large errors sometimes occur
- Relatively less sample will result in greater error

# 8. Conclusion

## 8.1 Innovations

- On the basis of endogenous economic growth model, we add the constraint of energy and environment and discuss the problem of sustainable economic growth under the constraints of energy and environment.

- The gray system forecast model is constructed to predict the total energy demand. Meanwhile, DEA model assessing supply indicators and the cleaner energy components index of the four states is built respectively.

- By comprehensively using the transcendental logarithm production function and the forecast result of this paper, the correlation between GDP, population and energy is analyzed and compared, and the future values forecasted and energy efficiency improvement plans are given.

## 8.2 Summary

Through the above work, we complete the data analysis of energy documents, constructing the DEA model to evaluate the energy structure of each state and giving a better description of the efficiency of different energy sources through different evaluation factors. Then we conclude that the performance of TX is the best. Then, we forecast the energy consumption in 2025 and 2050 through the gray forecast model. And the growth of cleaner renewable energy basically shows an exponential growth. Based on our criteria, we proposed the governors a viable target for cleaner renewable energy use, making our contribution to the solution of the energy problem by synthesizing various data and model predictions.

Appendix1: Average Price of Energy (million dollars per bbtu)

| AZ | CA | NM | TZ |
|---|---|---|---|
| 1.968725 | 1.734718 | 1.462613 | 1.29252 |
| 2.064549 | 1.810293 | 1.52972 | 1.338924 |
| 2.135317 | 1.909478 | 1.586197 | 1.416555 |
| 2.339663 | 2.12551 | 1.890197 | 1.487476 |
| 3.117465 | 3.03013 | 2.529958 | 2.276777 |
| 3.86824 | 3.463964 | 2.877919 | 2.819476 |
| 4.117297 | 3.799195 | 2.943338 | 3.165299 |
| 4.585677 | 4.261426 | 3.745225 | 3.506204 |
| 5.076325 | 4.566838 | 4.171871 | 3.707101 |
| 6.177505 | 5.527875 | 5.508216 | 4.755989 |
| 8.37574 | 7.707711 | 7.096522 | 6.051901 |
| 9.33057 | 8.844067 | 8.468766 | 7.122539 |
| 9.891835 | 9.289548 | 9.005497 | 7.590767 |
| 9.517988 | 8.840345 | 8.283514 | 7.535602 |
| 9.540809 | 8.737668 | 8.598074 | 7.479805 |
| 10.02399 | 8.890702 | 9.317712 | 7.069303 |
| 9.113931 | 7.74288 | 8.150882 | 5.778546 |
| 9.738058 | 7.631746 | 8.163816 | 5.764935 |
| 9.707741 | 7.831059 | 7.925141 | 5.598182 |
| 10.33209 | 8.247563 | 8.531832 | 5.636966 |
| 11.18754 | 8.990614 | 9.291936 | 6.463484 |
| 10.90462 | 9.100012 | 8.806349 | 6.329738 |
| 11.43518 | 9.584283 | 8.671833 | 6.271894 |
| 11.67808 | 9.52122 | 9.076703 | 6.473363 |
| 11.50009 | 9.56759 | 9.299665 | 6.630726 |
| 11.27115 | 9.64333 | 9.031393 | 6.447519 |
| 11.82933 | 9.946239 | 9.546554 | 7.049882 |
| 11.67528 | 10.32187 | 9.632373 | 6.897463 |
| 10.74631 | 9.288981 | 8.856925 | 6.132383 |
| 11.15313 | 9.984 | 9.007192 | 6.700013 |
| 12.86534 | 12.00892 | 10.8467 | 8.696622 |
| 12.77916 | 12.92874 | 11.34581 | 8.87571 |
| 12.35595 | 12.00603 | 10.57206 | 7.946983 |
| 13.83706 | 13.6799 | 11.90132 | 9.764047 |
| 15.31939 | 15.1167 | 13.51666 | 11.39398 |
| 17.8358 | 17.42068 | 16.52153 | 14.6886 |
| 19.6451 | 19.35315 | 18.33573 | 16.6676 |
| 20.72753 | 20.11723 | 19.0177 | 17.61111 |
| 23.75674 | 23.06592 | 22.4287 | 21.5495 |
| 19.66482 | 18.40528 | 17.17897 | 15.37925 |

Appendix2: Population Amount

|    | CA   | TX    | AZ    | NM    |
|----|------|-------|-------|-------|
| 60 | 1587 | 962.4 | 132.1 | 95.4  |
| 61 | 1650 | 982   | 140.7 | 96.5  |
| 62 | 1707 | 1005  | 147.1 | 97.9  |
| 63 | 1767 | 1016  | 152.1 | 98.9  |
| 64 | 1815 | 1027  | 155.6 | 100.6 |
| 65 | 1858 | 1038  | 158.4 | 101.2 |
| 66 | 1886 | 1049  | 161.4 | 100.7 |
| 67 | 1918 | 1060  | 164.6 | 100   |
| 68 | 1939 | 1082  | 168.2 | 99.4  |
| 69 | 1971 | 1104  | 173.7 | 101.1 |
| 70 | 1997 | 1120  | 177.5 | 101.7 |
| 71 | 2035 | 1151  | 189.6 | 105.4 |
| 72 | 2059 | 1176  | 200.8 | 107.9 |
| 73 | 2087 | 1202  | 212.4 | 100.6 |
| 74 | 2117 | 1227  | 222.3 | 113.1 |
| 75 | 2154 | 1257  | 228.5 | 116   |
| 76 | 2194 | 1290  | 234.6 | 118.9 |
| 77 | 2235 | 1319  | 242.5 | 121.6 |
| 78 | 2284 | 1350  | 251.5 | 123.8 |
| 79 | 2326 | 1389  | 263.6 | 128.5 |
| 80 | 2367 | 1423  | 271.7 | 130.3 |
| 81 | 2429 | 1475  | 281   | 133.3 |
| 82 | 2482 | 1533  | 289   | 136.4 |
| 83 | 2536 | 1575  | 296.9 | 139.4 |
| 84 | 2584 | 1601  | 306.7 | 141.7 |
| 85 | 2644 | 1627  | 318.4 | 143.8 |
| 86 | 2710 | 1656  | 330.8 | 146.3 |
| 87 | 2778 | 1662  | 343.7 | 147.9 |
| 88 | 2846 | 1667  | 353.5 | 149   |
| 89 | 2922 | 1681  | 362.2 | 150.4 |
| 90 | 2996 | 1706  | 368.4 | 152.2 |
| 91 | 3047 | 1740  | 378.9 | 155.5 |
| 92 | 3097 | 1776  | 391.6 | 159.5 |
| 93 | 3127 | 1816  | 406.5 | 163.6 |
| 94 | 3148 | 1856  | 424.5 | 168.2 |
| 95 | 3170 | 1896  | 443.2 | 172   |
| 96 | 3202 | 1934  | 485.7 | 175.2 |
| 97 | 3249 | 1974  | 473.3 | 177.5 |
| 98 | 3299 | 2016  | 488.3 | 179.3 |
| 99 | 3350 | 2056  | 502.4 | 180.8 |

Appendix3: Inflation Convert

| Year | Factor | Year | Factor |
|------|--------|------|--------|
| 1960 | 5.7221 |      |        |
| 1961 | 5.6596 | 2001 | 1.194  |
| 1962 | 5.591  | 2002 | 1.1759 |
| 1963 | 5.5285 | 2003 | 1.1529 |
| 1964 | 5.4448 | 2004 | 1.1221 |
| 1965 | 5.347  | 2005 | 1.0871 |
| 1966 | 5.201  | 2006 | 1.0547 |
| 1967 | 5.0541 | 2007 | 1.0274 |
| 1968 | 4.848  | 2008 | 1.0076 |
| 1969 | 4.6206 | 2009 | 1      |
| 1970 | 4.389  |      |        |
| 1971 | 4.1769 |      |        |
| 1972 | 4.0035 |      |        |
| 1973 | 3.7969 |      |        |
| 1974 | 3.484  |      |        |
| 1975 | 3.1887 |      |        |
| 1976 | 3.0227 |      |        |
| 1977 | 2.8462 |      |        |
| 1978 | 2.6594 |      |        |
| 1979 | 2.4566 |      |        |
| 1980 | 2.2534 |      |        |
| 1981 | 2.061  |      |        |
| 1982 | 1.9406 |      |        |
| 1983 | 1.8609 |      |        |
| 1984 | 1.8029 |      |        |
| 1985 | 1.747  |      |        |
| 1986 | 1.7125 |      |        |
| 1987 | 1.6699 |      |        |
| 1988 | 1.6134 |      |        |
| 1989 | 1.553  |      |        |
| 1990 | 1.4976 |      |        |
| 1991 | 1.4494 |      |        |
| 1992 | 1.4471 |      |        |
| 1993 | 1.3841 |      |        |
| 1994 | 1.3553 |      |        |
| 1995 | 1.3276 |      |        |
| 1996 | 1.3038 |      |        |
| 1997 | 1.2819 |      |        |
| 1998 | 1.2681 |      |        |
| 1999 | 1.249  |      |        |
| 2000 | 1.2212 |      |        |